\documentclass[11pt,twoside]{atmp}
\usepackage{amsmath,amssymb}

\usepackage[all]{xy}
\usepackage{color}

\usepackage{bm}
\def\R{\mathbb{R}}
\def\C{\mathbb{C}}

\def\Z{\mathbb{Z}}
\def\sgn{{\rm sgn}}
\newcommand{\eps}{\varepsilon}

\newcommand{\rechts}{\rightarrow}

\newcommand{\Seff}{S_{\text{eff}}}
\newcommand{\tr}{\text{Tr}}

\theoremstyle{definition}
\newtheorem*{Acknow}{Acknowledgement}
\newtheorem*{Conclusion}{Conclusion}
\begin{document}

\title%[Zero-Field Hall Effect in (2+1)-dimensional QED]
{Zero-Field Hall Effect in (2+1)-dimensional QED}

%\arxurl{cond-mat/0505428}

%\preprint{DIAS-STP-05-14}
\begin{flushright}
\begin{tabular}{l}
DIAS-STP-05-14 
%cond-mat/0505428
%December 2008
\end{tabular}
\end{flushright}

\author[Marianne Leitner]{Marianne Leitner}

\address{Dublin Institute for Advanced Studies\\
School of Theoretical Physics \\
10 Burlington Road, Dublin 4, Ireland}  
%lines should be separated with double backslashes: \\
\addressemail{leitner@stp.dias.ie}

\begin{abstract}
In QED of two space dimensions, a quantum Hall effect occurs in the absence of any magnetic field.
We give a simple and transparent explanation.
In solid-state physics, the Hall conductivity for non-degenerate ground state is expected to be given by an integer, the Chern number. In our field-free situation, however, the conductivity is $\pm 1/2$ in natural units. We fit this half-integral result into the topological setting and give a geometric explanation reconciling the points of view of QFT and solid state physics. For quasi-periodic boundary conditions, we calculate the finite size correction to the Hall conductivity. Applications to graphene and similar materials are discussed.
\end{abstract}

\maketitle

\section{Introduction}

The Hall effect of quantum electrodynamics (QED) in two space dimensions occurs in the absence of any magnetic field.
The corresponding off-diagonal conductivity is $\pm 1/2$ in natural units. Here we present a geometric interpretation and calculate finite size corrections. Moreover, we discuss applications to graphene and related materials.\\
The discussion of this effect goes back at least to  
the articles by Redlich \cite{Redlich:1983} and Jackiw \cite{Jackiw:1984}, which focused on massless non-abelian gauge theory in $2+1$ dimensions, but included discussions of QED with a mass term. In principle, the Hall conductivity is implied by these calculations, though in (\cite{Redlich:1984}, formula (3.8)), this is obscured by an apparent factor of $2$ change in the normalisation of the Chern-Simons term.
The half-integrality of the correct result required the resolution of a paradox, since general arguments suggest integral values \cite{TKNN:1982, Haldane:1988}. 
It was stated that when spin is included, the conductivity is doubled \cite{Jackiw:1984}, 
but in certain situations this is no longer true when a Zeeman term is included \cite{Haldane:1988}
such that integrality is not necessarily restored.
The result for zero magnetic field has been stated in \cite{Ludwig:1994},
where a system is shown to undergo an integer quantum Hall transition as a control parameter is varied, in absence of disorder. The derivation of the Hall conductivity itself is purely computational, however, and does not give any insight into the half integral nature of the result. A straightforward geometrical interpretation of the fractional value should prevent errors by factors of $2$ in quantum  Hall effect (QHE) calculations and facilitate the analysis of more complicated systems.\\
The actual shift by $1/2$ from integral values has gained topicality by recent work on the electric properties of the new material graphene, when exposed to a strong external magnetic field \cite{Gusynin:2005, nature:2005}.
For low energy excitations, this monolayer of graphite realises a relativistic QFT in $2+1$ dimensions (with an analogue of the velocity of light). The Hall effect in graphene has been referred to as a half-integer effect (\cite{nature:2005}, p.\ 198, p.\ 201).\\
For calculations of the QHE, Feynman diagrams with two and more loops are irrelevant \cite{Coleman-Hill, KS:1985}.
One may consider a single fermion loop with two external photon lines \cite{AHW:1982} or, equivalently, the quantum mechanics of relativistic fermions in a constant external field \cite{Redlich:1984}.
We first consider the latter approach. The massive ($m\not=0$) Dirac equation reads 
\begin{equation*}
[-i(\partial_{\mu}+ieA_{\mu})\gamma^{\mu}+m]\psi=0,
\end{equation*}
where $e$ is the electron charge and $\bm{A}$ denotes an abelian external electromagnetic field.
We use the convention $\gamma^{\mu}:=\sigma_3\sigma_{\mu}$, for $\mu=0,1,2$, where $\sigma_{\mu}$ is the $\mu$-th  Pauli matrix, and 
$\{\gamma^{\mu},\gamma^{\nu}\}=2g^{\mu\nu}$ for $g_{\mu\nu}=diag(1,-1,-1)$. 
In a homogeneous background field  $F^{\mu\nu}\equiv\partial^{\mu}A^{\nu}-\partial^{\nu}A^{\mu}$, 
the ground state current is \cite{AHW:1982, Redlich:1984}  
\begin{equation}\label{AHW}
\langle j_{\mu}\rangle
=\frac{1}{8\pi}\:sgn(m)\:\eps_{\mu\nu\eta}(eF^{\nu\eta}).
\end{equation}
When $\mu=1,2$, this equation becomes the Ohm-Hall law with Hall conductivity
$\sigma_H:=\sigma_{21}h/e^2=\frac{1}{2}sgn(m)$ \cite{Semenoff:1984}.\\ 
This result is remarkable for two reasons. 
Firstly, the Hall conductivity is independent of the magnetic field strength $F^{12}$ and reveals a \emph{zero-field Hall effect} \cite{Froehlich:1991} for $F^{\mu\nu}=0$. Moreover, $\sigma_H$ is not an integer.
At first glance, this seems to contradict a basic result of solid state physics. It says that non-degeneracy of the ground state is the decisive condition for integrality, which is fulfilled for a relativistic electron gas without interactions between the electrons. Thus a closer look is needed to cast our half-integer result into the classical setting.\\
In the first section of this paper we explain why a thorough investigation of the Dirac sea gives acces to a better understanding of electronics in graphene and related materials. We then review the integer quantum Hall effect (Sect.\ \ref{Infinity problem}) and reformulate the Kubo-formula in the context of QED in two space dimensions (Sect.\ \ref{QED}). We shall see that for our system the old time-ordered perturbation methods are equivalent to the relativistically covariant ones, but more efficient.  This is very useful for systems with boundaries, where Lorentz invariance does anyhow not apply \cite{GruLei:2004}. In Sect.\ \ref{Geometry}, we calculate and discuss half-integral value of $\sigma_H$. Though the standard topological description fails, $\sigma_H$ has a nice geometric interpretation. As a side product, we calculate the corrections to $\sigma_H$ in a finite area with periodic boundary conditions.  

\section{The effective low energy theory of graphene and similar materials}

From the point of view of solid state physics the zero field Hall effect is uncommon.
Since by (\ref{AHW}), $\sigma_H$ changes sign under space or time reflexion, one indeed expects $\sigma_H=0$ under many circumstances.
To break time-reversal invariance, usually a magnetic field perpendicular to the plane is incorporated into the Hamiltonian to obtain a nonzero $\sigma_H$ \cite{Asshap:1986}. The mechanism to produce a QHE through breaking of time-reversal symmetry has been explored in more detail by \cite{Haldane:1988}. The discussion is based on the work of Semenoff \cite{Semenoff:1984} who suggested a planar honeycomb configuration as a solid state analogue of $(2+1)$-dimensional QED \cite{Redlich:1984}. Here there are two inequivalent atomic species, corresponding to two underlying triangular lattices, per fundamental cell. For each of them, the continuum limit of a (nearest-neighbour) tight-binding approximation yields a nonmagnetic Dirac operator of the type we are going to consider here. 
In other words, for low energy excitations, the charge carriers in this carbon monolayer can be treated as relativistic massless particles in $2+1$ dimensions (with an analogue of the velocity of light).
Experiments on graphene realise part of Semenoff's expectation, in particular the Lorentz invariance of the low energy behaviour \cite{KNG}. 
Due to fermion doubling (touching of conductance and valence band at two points, $K$ and $K'$, which correspond to Bloch waves supported on either sublattice), it is invariant under both parity and inversion (rotation by $\pi$).
The presence of a mass breaks time-reversal invariance of the \emph{individual} Hamiltonians, but the mass inversion can be compensated by an exchange of $K$ and $K'$. In particular, the respective contributions to the net Hall current cancel each other.
In agreement with this argument, graphene by itself is symmetric under space and time reflections, so that a strong magnetic field is indispensible for a Hall effect.
Haldane implements a local magnetic flux density orthogonally to the plane in such a way that only the second-neighbour hopping terms are affected, which he also includes. Now time-reversal symmetry of the model is manifestly broken provided the respective effective masses for $K$ and $K'$ are of the same sign (or of opposite sign in Haldane's rather artificial conventions). In this case $\sigma_H\not=0$ is obtained. This is true when the spectrum is discretised by an external magnetic field, but also for continuous spectrum in the zero field case.\\
The new hopping terms introduce an additional parameter in the model by which the masses can be varied independently. This defeats fermion doubling: The effective low energy theory needs not contain two relativistic fermions at all. Instead, a heavy mass fermion (relativistic or not) may become a spectator, which can be integrated out, leaving only a Chern-Simons contribution at the low energy \cite{Haldane:1988, Ludwig:1994,HKW:1996}. The important fact for us is that this term produces the same Hall conductivity \cite{Ishikawa:1984}, so that a value of $\pm 1/2$ will eventually fit into the usual integer quantum Hall effect setting.\\ 
The specific energy dispersion away from $K$ and $K'$ is part of high energy physics and does not affect our discussion (apart from the Chern Simons term). 
Thus a thorough study of a single fermion and possibly a Chern-Simons term are important steps towards an understanding of the Hall effect in graphene and modifications thereof.\\
Experiments show that the first Hall plateau (starting from charge carrier concentration zero) appears already at half the normal filling, whereas each higher plateau is reached from there by a ladder of integer steps.
\cite{Gusynin:2005} tries to explain this unconventional behaviour by the fact that the $E=0$ level in graphene is shared by electrons and holes, whereas every other electron Landau level bears an antiparticle's counterpart. This statement leads \cite{nature:2005} to conclude that the shift from integer valued plateaux by $1/2$ is lifted when a gap opens around the cone singularity, and to originate the difference from the conventional QHE in ordinary graphite in the fact that here the charge carriers have a nonzero effective mass. The reasoning is not stringent. We will actually see that presence of a mass is not the crucial point.\\  
In Haldane's model, the net magnetic flux per unit cell is zero (the magnetic field doesn't affect nearest-neighbour hopping terms). A consequent step further is to forgo any external magnetic field, since a non-symmetric modification of graphene with some interior magnetic field is conceivable \cite{Semenoff:1984,Khv:2001}. For applications in computer technology where a large number of components has to be put in place inside of small volumes, this would allow much greater flexibility than graphene.
%\cutpage %move this line so that the first page breaks at the appropriate place.

%\setcounter{page}{<insert page # for second page>}

\noindent
% (d.h. kein Einruecken am Anfang eines Abschnitts)

\section{The Kubo formula in the integer quantum Hall effect}\label{Infinity problem}

We review the general argument for the integrality of the Hall conductivity. 
$\sigma_H(\vec{x})$ is derived by perturbation theory.
To first order one has
\begin{equation}\label{Stoerungstheorie}
\delta\langle 0|j_{\mu}(\vec{x})|0\rangle
=\langle 0|j_{\mu}(\vec{x})\delta|0\rangle+c.c.,
\end{equation}
where $|0\rangle$ denotes the ground state.
For
\begin{equation}\label{perturbation operator}
\delta H
:=\int A_0(\vec{x})\rho(\vec{x})\:d^2x
\end{equation}
where $A_0(\vec{x})=-E_ix_i$ and $\rho$ is the charge density, variation of (\ref{Stoerungstheorie}) w.r.t.\ the electric field  gives the Kubo-formula for the Hall-conductivity at zero temperature.\\
(\ref{perturbation operator}) applies to infinite volume, but mesoscopic devices are better approximated by finite systems with periodic boundary conditions \cite{AS:1985}. Thus we consider the torus $T=\R^2/\Lambda$ for some finite lattice $\Lambda\cong\Z^2$, and with area ${\mathcal{A}}$ of the fundamental cell. In this case, $x_i$ is only locally a well-defined coordinate. The description of the system needs additional parameters $(k_1,k_2)$. If one uses $A_i=E_it$ (with $t\in\R$) to define the electric field, the parameters describe background fluxes $eA_i=k_i$. 
Locally they can be compensated by gauge transformations $\psi(\vec{x})=\exp[i\vec{k}\cdot\vec{x}]\:u(\vec{x})$ (for one-particle wave functions), but globally this only works for $\vec{K}\in\Lambda^*$, where $\Lambda^*:=\{\vec{K}\in\R^2\mid\vec{K}\cdot\vec{R}\in 2\pi\Z,\:\forall\vec{R}\in\Lambda\}$ is the dual lattice.
For time-independent $\delta H$ and $P^{(0)}\equiv|0\rangle\langle 0|$,
\begin{equation*}
\delta|0\rangle=(E_0-H)^{-1}[P^{(0)}]^{\perp}\delta H|0\rangle.
\end{equation*}
Otherwise, the solution of the time-dependent Schr\" odinger equation is given by
 $|\psi\rangle
=\exp[-iE(\vec{k})t]\:(|0\rangle+\delta|0\rangle)$, where
\begin{equation*}
\delta|0\rangle=i(E_0-H)^{-2}\left(E_i\int j_i\:d^2x\right)|0\rangle,
\end{equation*}
since $\delta H=-\int j_i\:\delta A_i\:d^2x$.
This yields the Kubo-formula for the ground state's contribution to the Hall conductivity (with $\hbar\dot{=}1$)  
\begin{equation}\label{curvature}
\sigma_{21}^{(0)}(\vec{k})
=\frac{ie^2}{{\mathcal{A}}(T)}\text{Tr}\left(P^{(0)}_{\vec{k}}\left[\partial_{k_1}P^{(0)}_{\vec{k}},\partial_{k_2}P^{(0)}_{\vec{k}}\right]\right).
\end{equation}
Here $P^{(0)}_{\vec{k}}$ is the projector onto the ground state satisfying the boundary condition $\vec{k}\in T^*$, the dual torus.
Note that the Hall conductivity (\ref{curvature}) depends on time. For $A_i=E_it$, the parameter $\vec{k}$ follows a geodesic in $T^*$. For rational slope, we predict a periodic variation of the Hall conductivity, similar to the AC Josephson effect. Otherwise one finds of course quasi-periodic behaviour.\\
We assume, as for the rest of this paper, that the Fermi energy lies in a spectral gap and that the temperature is zero.
When a periodic potential with period lattice $\Gamma$ splits the spectrum into energy bands, then the electron states of the $n$-th band form a line bundle over the torus $T^*=\R^2/\Gamma^*$, provided no energy degeneracy occurs. Let $P^{(n)}_{\vec{k}}$ denote the projection onto the $n$-th band state. Then the bundle is equipped with the natural (adiabatic) connection $P^{(n)}_{\vec{k}}\circ\nabla_{\vec{k}}$ whose curvature-form is defined through (\ref{curvature}), with $P^{(0)}_{\vec{k}}$ replaced by $P^{(n)}_{\vec{k}}$.
In particular, the average of $\sigma_{21}^{(n)}(\vec{k})$ over the boundary conditions yields, using ${\mathcal{A}}(T){\mathcal{A}}(T^*)=(2\pi)^2$, the Chern number 
\begin{equation*}
\sigma_H^{(n)}
=\frac{1}{{\mathcal{A}}(T^*)}\int_{T^*}\sigma_H^{(n)}(\vec{k})\:d^2k
\end{equation*}
of this bundle \cite{TKNN:1982,Kohmoto:1985}, which is an integer.
In general, the states under consideration are multi-particle states with a macroscopic but finite number of particles.
When the system is a macroscopic torus $\R^2/\Lambda$, the ground states which vary through their quasi-periodic boundary conditions form a line bundle over $\R^2/\Lambda^*$, if no degeneracies occur.
Then (\ref{curvature}) applies to the multi-particle projector $P^{(0)}_{\vec{k}}$, and the Chern number of this bundle yields the (averaged) Hall conductivity \cite{Niu-Thou:1985, AS:1985}.
When the lattice $\Lambda$ is macroscopic, finite size effects are irrelevant and the averaging is trivial.\\ 
In many cases, the ground state can be approximated by a finite wedge product of eigenstates of single-particle Hamiltonians. This is very convenient because then all manipulations of the ground state reduce to one-particle computations. In particular, the multi-particle trace (\ref{curvature}) splits into a finite sum of traces in low dimensions. Thus the Hall effect in this setting features essentially just one-particle physics.

\section{The Kubo formula in QED$_3$}\label{QED}

As we have seen, $\sigma_H(\bm{x})\equiv\sigma_H(t,\vec{x})$ (for fixed time $t$) can be derived by \textquotedblleft old-fashioned\textquotedblright perturbation theory in the Schr\"odinger picture. For more general problems in QED$_3$, e.g.\ non-homogeneous external fields, it is instructive to relate the resulting Kubo formula to the standard relativistically invariant treatment. Rewrite the r.h.s.\ of eq.\ (\ref{Stoerungstheorie}) as an integral over the entire space-time.
The unperturbed Dirac operator does not depend on time, 
and changing from Schr\" odinger into Heisenberg picture,
\begin{align}
\delta\langle 0|j_{\mu}(\bm{x})|0\rangle
&=
-i
\int_{-\infty}^{\infty}
\langle 0|T\left[j_{\mu}(\bm{x})(P_{|0\rangle})^{\perp}\delta H(t)\right]|0\rangle 
\:dt\nonumber\\
&=-i\delta e
\left(\int_{\R^3}
\langle 0|T[j_{\mu}(\bm{x})\mathcal{O}(\bm{x}')]|0\rangle
\:d^3x'\right)_{\text{regul.}}\nonumber\\
\label{relativistic invariant Kubo}
&=
\delta e\left(\int_{\R^3}
\langle 0|j_{\mu}(\bm{x})\mathcal{O}(\bm{x}')|0\rangle_{\text{Euclid.}}
\:d^3x'\right)_{\text{regul.}},
\end{align}
using Wick rotation in (\ref{relativistic invariant Kubo}).
Here $\mathcal{O}(\bm{x})=A^{\nu}(\bm{x})j_{\nu}(\bm{x})$
is the multi-particle operator with $A^0(\bm{x})=-ex_iE^i$ and $A^j\equiv const.$ for $j=1,2$.
(\ref{relativistic invariant Kubo}) is the relativistically invariant version of the quantum mechanical Kubo formula in three-dimensional QED. 
To obtain the Hall conductivity, 
one needs \cite[form.\ (2.5)]{AHW:1982}
\begin{equation*}
\left[
\langle0|j^{\mu}(\bm{x})j^{\nu}(\bm{x}')|0\rangle_{\text{Euclid.}}
\right]_{\text{regul.}}
\sim
\sgn(m)\:\eps^{\mu\eta\nu}\frac{\partial}{\partial x^{\eta}}\:\delta^{(3)}(\bm{x}-\bm{x}').
\end{equation*}
Note that the $\sgn(m)$ factor on the r.h.s.\ is necessary for Lorentz invariance, since a change in space-time orientation can be compensated by a sign change of $m$.
Using  
\begin{equation}\label{tangent vector}
\langle j^{\mu}(\bm{x})\rangle
:=\frac{\delta\Seff[\bm{A}]}{e\delta A_{\mu}(\bm{x})}\:,
\end{equation}
the regularised effective action turns out to be, to the order $O(e^3)$, the Chern-Simons action \cite{Redlich:1984}
\begin{equation*}
S_{CS}[\bm{A}]
=\sgn(m)\frac{e^2}{8\pi}\eps^{\mu\nu\eta}\int_{\R^3}(\partial_{\mu}A_{\nu})A_{\eta}\:d^3x.
\end{equation*}
This proves (\ref{AHW}).
In virtue of (\ref{tangent vector}), $\langle j^{\mu}(\bm{x})\rangle$ is a tangent vector on the moduli space of (regularised) quantum field theories in $2+1$ dimensions. On the basis of our discussion in Sect.\ \ref{Infinity problem}, the corresponding two-point function 
\begin{equation}\label{contact term}
\langle j^{\mu}(\bm{x})j^{\nu}(\bm{x}')\rangle_{\text{Euclid.}}
=
\frac{1}{8\pi}\sgn(m)\:\eps^{\mu\nu\eta}\frac{\partial}{\partial x^{\eta}}\:\delta^{(3)}(\bm{x}-\bm{x}')+....
\end{equation}
should be compared to the operator product expansion (OPE) of two tangent vectors on the moduli space two-dimensional conformal field theories (CFT) \cite[form.\ (2)]{Kutasov}. In this OPE, the local contact term contains the connection on the tangent bundle. Similarly, from (\ref{contact term}) we can read off the curvature of the family of vacua parametrised by $T^*$. 
This family can however not be interpreted as a line bundle, as we will see in the following section.

\section{Geometry of the zero-field Hall effect}\label{Geometry}

The example of interest to us is the constant Dirac operator
\begin{equation*}
H
:=-i\vec{\nabla}\cdot\vec{\sigma}+m\sigma_3,
\end{equation*}
acting on the smooth functions in ${\mathcal{H}}:=L^2(\R^2,\C)\otimes\C^2$.
Time inversion is implemented by the anti-unitary operator $U_T:{\mathcal{H}}\rechts {\mathcal{H}}$, defined by
\begin{equation*}
U_T(\psi):=\sigma_2\bar{\psi}.
\end{equation*}
(Here $\bar{\psi}$ is the complex conjugate of $\psi$.) The mass term breaks time reversal symmetry: 
$U_T\circ H\circ U_T^{-1}=-i\vec{\nabla}\cdot\vec{\sigma}-\sigma_3m\not=H$.
To calculate the Hall conductivity for independent electrons in a mesoscopic volume,
we consider the direct integral decomposition of ${\mathcal{H}}$ w.r.t.\ the representation of the translation group $\Lambda$. 
${\mathcal{H}}':= L^2(T,\C)\otimes\C^2$ is acted upon by $H(\vec{k})$, the conjugate of $H$ by $e^{-i\vec{k}\cdot\vec{x}}$.
Introducing quasi-periodic boundary conditions on $\R^2$ results in the superposition of the energy hyperboloid with any of its $\Lambda^*$-translates. $H(\vec{k})$ has the discrete spectrum
\begin{equation*}
E^{(\pm)}_{\vec{K}}(\vec{k})=\pm[(\vec{k}+\vec{K})^2+m^2]^{1/2}
\end{equation*}
for $\vec{K}\in\Lambda^*$. The Hall conductivity is easily obtained by use of Fourier transformation, which maps ${\mathcal{H}}'$ onto $\ell^2(\vec{k}+\Lambda^*)\otimes\C^2$, transforming $H(\vec{k})$ into
$\oplus_{\vec{K}\in\Lambda^*}((\vec{k}+\vec{K})\cdot\vec{\sigma}+m\sigma_3)$. 
Let $P^{(-)}_{\vec{k}+\vec{K}}$ denote the spectral projector to the energy $E^{(-)}_{\vec{K}}(\vec{k})$.
The contribution of the negative energies to the Hall conductivity is
\begin{align}
\label{Trace in Fourier space}
\sigma_{21}^{(-)}(\vec{k})
&=\frac{ie^2}{{\mathcal{A}}(T)}\sum_{\vec{K}\in \Lambda^*}\tr_{\C^2}\left(P^{(-)}_{\vec{k}+\vec{K}}\left[\partial_{k_1} P^{(-)}_{\vec{k}+\vec{K}},\partial_{k_2}P^{(-)}_{\vec{k}+\vec{K}}
\right]\right)\\
\label{sigma(k)}
&=\frac{e^2}{2\:{\mathcal{A}}(T)}
\sum_{\vec{K}\in\Lambda^*}\frac{m}{[(\vec{k}+\vec{K})^2+m^2]^{3/2}}.
\end{align}
Similar finite-size corrections to the Hall conductivity have been considered in \cite{ASZ:1994}, but there the effects vanish for translationally invariant systems on a torus.
Using Poisson summation, we obtain from (\ref{sigma(k)}),
\begin{equation*}
\sigma_{21}^{(-)}(\vec{k})
=\frac{e^2}{4\pi}\:\sgn(m)\sum_{\vec{R}\in\Lambda}
e^{-|m||\vec{R}|-i\vec{k}\cdot\vec{R}}.
\end{equation*}
In the limit $\Lambda\rechts\infty$ (i.e., all space lattice periods large) we obtain the Hall conductivity on the infinite plane,
\begin{equation}\label{sigma-H}
\sigma_H^{(-)}
=\frac{1}{2}\:\sgn(m).
\end{equation}
It is instructive and important for generalisation to reconcile this half-integer result with the arguments for integral Hall conductivity in solid state physics.
The bundle description assumes the presence of a single state for a fixed band index and each $\vec{k}\in T^*$.
When finitely many degeneracies occur, one may consider a finite cover of the torus $\R^2/\Lambda^*$ to save this argument.
The present system bears a discrete symmetry by which the intersections occur pairwise and are superposed. However, every hyperboloid will for high enough energy (cones far apart on $\R^2$) intersect every other hyperboloid. Thus over every degeneracy point there is  an infinite number of energy crossings. But the infinite cover $\R^2$ is non-compact. In particular, the index theorem does not apply, and the Hall conductivity need not be integral.
On the other hand, in the presence of a magnetic field, the Chern number is one for each individual Landau level. Indeed, the contributions of the infinitely many bands sum up to
\begin{equation*}
\zeta(0)
=\sum_{n=1}^{\infty}n^0
=-1/2
\end{equation*}
or to $+1/2$, if the $n=0$ level is included. Eventually, as we have seen, the Hall conductivity in this system is independent of the field strength. The use of the $\zeta$-function regularisation would not be convincing without further arguments, but is analogous to the derivation of the critical dimension in string theory, based on $\zeta(1)=-1/12$ \cite{BN:1973}.\\
The multi-particle formalism shows up in form.\ (\ref{Trace in Fourier space}), but the wedge products to consider exploit the infinite depth of the Dirac sea. Though in QED, non-degeneracy of the ground state is assured by the Pauli exclusion principle, the Hall conductivity is not an integer any more. The reason is that the Dirac sea does not define a line bundle over $T^*$.
Let $e^{\pm}_{\vec{k}+\vec{K}}\in L^2(T,\C)\otimes\C^2$ be the eigenvector of $H(\vec{k})$ to the energy $E^{\pm}_{\vec{K}}(\vec{k})$. Then
\begin{equation*}
|0(\vec{k})\rangle:=\wedge_{\vec{K}\in\Lambda^*}e^{-}_{\vec{k}+\vec{K}}
\end{equation*}
is the Dirac sea. Here, $\Lambda^*$ be ordered by energy, i.e., by the distance of $\vec{k}\in\R^2$ from the respective cone centers $\vec{K}\in\Lambda^*$. Each energy degeneracy induces an interchange of two neighbouring wedge factors as $|0(\vec{k})\rangle$ passes through a degeneracy point, but their infinite number makes it impossible to determine the overall sign change. Note that this problem occurs already locally, because the set of degeneracy points - lines in $\R^2$ orthogonally bisecting shortest paths connecting any two points in $\Lambda^*$ - lies dense.\\
To dispose of this somewhat artificial ordering issue, redefine the Dirac sea $|0(\vec{k})\rangle$ as
\begin{equation*}
\otimes_{\vec{K}\in\Lambda^*}e^-_{\vec{k}+\vec{K}}
\end{equation*}
(for any fixed order on $\Lambda^*$). This approach yields isomorphic product spaces, since the $e^{\pm}_{\vec{k}+\vec{K}}$ to different $\vec{K}\in\Lambda^*$ are orthogonal.
A suitable Hilbert space ${\mathcal H}(\vec{k})$ has the basis $\otimes_{\vec{K}\in\Lambda^*}e^{\epsilon(\vec{K})}_{\vec{k}+\vec{K}}$,
where $\epsilon(\vec{K})\in\{-1,1\}$ and equals $-1$ for almost all $\vec{K}\in\Lambda^*$.
We calculated the scalar product $\langle 0(\vec{k})|0(\vec{k}')\rangle$. Note that this is the same for wedge products, because only the trivial permutation contributes. The result diverges to zero (like an inverse power) for $\vec{k}\not=\vec{k}'$. Hence vaccum vectors to different parameters are not elements of one overall Hilbert space, and therefore we do not obtain an embedded line bundle over $T^*$.
Remarkably, the phase of the scalar product is still well-defined. Thus while the projector $P^{(-)}_{\vec{k}}$ onto $|0(\vec{k})\rangle$ must be treated formally in the connection $P^{(-)}_{\vec{k}}\circ\nabla_{\vec{k}}$ on the family of vacua, its curvature makes good sense in the limit $\vec{k}'\rightarrow\vec{k}$.\\
The result (\ref{sigma-H}) is of geometric origin:
Another way to obtain it is by averaging (\ref{sigma(k)}) over the boundary conditions, 
\begin{equation*}
\sigma_H^{(-)}
=\frac{1}{{\mathcal{A}}(T^*)}\int_{T^*}\sigma_H^{(-)}(\vec{k})\:d^2k
=\frac{1}{4\pi}\int_{\R^2}\frac{m}{[\vec{k}^2+m^2]^{3/2}}\:d^2k.
\end{equation*}
The form of the Hamiltonian in momentum space $H(\bm{k})=\bm{k}\cdot\bm{\sigma}$ with $\bm{k}=(k_1,k_2,m)$ and $\bm{\sigma}:=(\sigma_j)_{j=1}^3$ suggests to embed $\R^2$ into $\R^3$, with image given by
$F_m:=\{\bm{k}\in\R^3|k_3=m\}$. The integrand
\begin{equation*}
\frac{m}{[\vec{k}^2+m^2]^{3/2}}\:dk_1\wedge dk_2
\end{equation*}
on $\R^2$ is the pull-back of the two-form
\begin{equation*}
\eta
:=\frac{1}{2}\:\eps^{\alpha\beta\gamma}
\:\frac{k_{\alpha}dk_{\beta}\wedge dk_{\gamma}}{|\bm{k}|^3}\:.
\end{equation*}
Since $\eta$ is invariant under the transformation $\bm{k}\mapsto\lambda(\bm{k})\bm{k}$, we can project $F_m$ onto the upper or lower half $S_{\pm}^2$ of the unit sphere $S^2\subset\R^3$, depending on the sign of $m$. On the unit sphere, $\eta$ restricts to the volume form times $\sgn(m)$. This gives 
\begin{equation*}
\sigma_H^{(-)}
=\frac{1}{4\pi}\int_{S^2_{\pm}\subset\R^3}\eta\:
=\:\frac{1}{2}\:\sgn(m)
\end{equation*}
geometrically, describing $\sigma_H^{(-)}$ as a solid angle.
The half-spheres are orbits of the Lorentz group,
so that the quantisation of $\sigma_H^{(-)}$ follows from Lorentz invariance.
This derivation shows that $\sigma_H$ can in principle take any value. It suffices to make a homeomorphism precede the projection onto $S^2$, which has the effect of shrinking the solid angle while it of course breaks Lorentz invariance, too. In our example, $\sigma_H$ is the shape of the surface in momentum space rather than a bundle topological quantity.\\
We have approached the \emph{zero-field Hall effect} in two ways.
In QED, we considered the perturbation of the vacuum current by $\mathcal{O}(\bm{x})$.
This approach requires a regularisation (which breaks parity but preserves gauge invariance). 
Alternatively, all computations can be restricted a priori to finite volume. This amounts to perturbing the low energy Lagrangian by $\vec{k}\cdot\vec{j}(\bm{x})$, 
where $\vec{k}=(k_1,k_2)\in T^*$ are the Aharonov Bohm fluxes through the fundamental cycles of $T=\R^2/\Z^2$. 
Thus the $\vec{k}\in T^*$ define non-equivalent (Lorentz non-invariant) quantum field theories in $2+1$ dimensions, over which we eventually average.

\begin{Acknow}
The author is grateful to R.\ Seiler and G.M.\ Graf for helpful discussions and would like to thank W.\ Nahm for support in the late stages of this work.
\end{Acknow}

\begin{Conclusion}
The Hall effect for relativistic massive fermions is important both for quantum field theory and condensed matter physics. It can be treated by the same formalism in both contexts,
which allows to visualise both the close analogies and the profound differences of the two physical systems. The corresponding half integral value of the Hall conductivity has an elegant geometric interpretation. Integrality breaks down since the quantum field theory does not allow for the existence of a global Hilbert space which is independent of the boundary conditions. This is analogous to the situation in Haag's theorem (for Lorentz invariant theories): There is no natural way to identify the Hilbert spaces of quantum field theories with different physical parameters. The boundary condition encodes a time dependence of the Hall conductivity, which to observe in experiments would be an interesting task.
Currently, relevant experimental results are restricted to graphene, where reflection invariance requires the use of strong external magnetic fields. In suitable non-symmetric materials, however, the effect may not need such a field.

\end{Conclusion}
%\begin{appendix}

%<insert any appendices here>

%\section{<appendix section title>}
%<insert appendix section text>

%\end{appendix}

\bibliographystyle{my-h-elsevier}

\end{document}